\documentclass[preprint,12pt]{elsarticle}

\usepackage{amssymb}
\usepackage{amsmath}
\usepackage{color}
\usepackage[version=4]{mhchem}

\journal{Current Opinion in Chemical Engineering}

\begin{document}

\begin{frontmatter}



\title{Prospects for Using Artificial Intelligence to Understand Intrinsic Kinetics of Heterogeneous Catalytic Reactions}


\author[GT]{Andrew J. Medford} 
\author[GT]{Todd N. Whittaker}
\author[GT]{Bjarne Kreitz}
\author[GT]{David W. Flaherty}
\author[CMU]{John R. Kitchin}

\affiliation[GT]{organization={School of Chemical \& Biomolecular Engineering, Georgia Institute of Technology},
            addressline={311 Ferst Drive NW}, 
            city={Atlanta},
            postcode={30332}, 
            state={GA},
            country={USA}}
\affiliation[CMU]{organization={Department of Chemical Engineering, Carnegie Mellon University},
            addressline={5000 Forbes Street}, 
            city={Pittsburgh},
            postcode={15213}, 
            state={PA},
            country={USA}}

\begin{abstract}
Artificial intelligence (AI) is influencing heterogeneous catalysis research by accelerating simulations and materials discovery. A key frontier is integrating AI with multiscale models and multimodal experiments to address the ``many-to-one'' challenge of linking intrinsic kinetics to observables. Advances in machine-learned force fields, microkinetics, and reactor modeling enable rapid exploration of chemical spaces, while \textit{operando} and transient data provide unprecedented insight. Yet, inconsistent data quality and model complexity limit mechanistic discovery. Generative and agentic AI can automate model generation, quantify uncertainty, and couple theory with experiment, realizing ``self-driving models'' that produce interpretable, reproducible, and transferable understanding of catalytic systems.
\end{abstract}

\end{frontmatter}














\section{Introduction}
\label{sec:intro}

The concept of artificial intelligence (AI) in heterogeneous catalysis is widely discussed but has no clear definition. One well-established example of AI in catalysis is the use of ``foundational'' or ``universal'' models of quantum chemistry, which have been widely applied for catalyst screening and are increasingly ubiquitous in computational catalysis workflows \cite{Chanussot2021, Tang2024}; however, they are not specific to catalysis. Numerous other examples of AI or machine learning (ML) in catalysis exist, ranging from reaction network analysis, spectroscopic analysis, data extraction from the literature, high-throughput screening, and, recently, self-driving smart labs \cite{Xin2025}. Most of these studies are undoubtedly examples of ``AI in catalysis,'' but there is substantial room for growth. In particular, there are few, if any, examples of AI models designed to address the full multiscale nature of catalysis, spanning from electrons to reactors, or the reality that catalysis science often involves piecing together incomplete or uncertain data. A key goal in catalysis science is to reveal a consistent picture of the active site and the reaction mechanism, but most existing AI models are not structured to yield detailed understanding of specific catalytic systems. Currently, chemical intuition drives mechanistic hypotheses and evaluations, but this is limited both in speed and scope. AI tools have the potential to accelerate the process and reduce bias by broadening mechanistic searches to cover a larger chemical reaction space and follow more reproducible procedures.

\begin{figure}[htbp]
  \centering
  \includegraphics[width=0.7\textwidth]{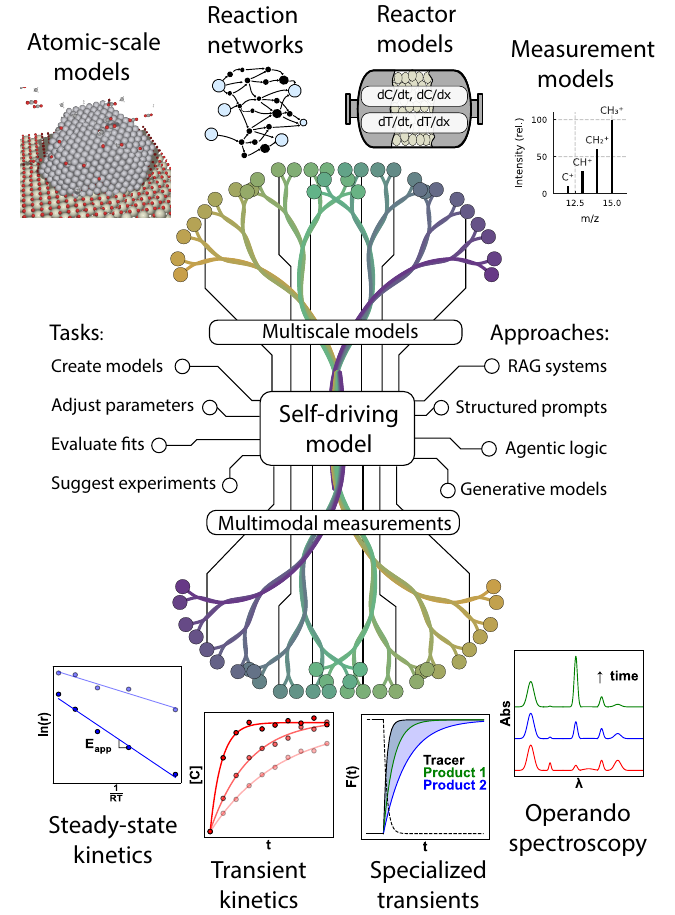}
  \caption{Schematic illustration of how a self-driving model connects multiscale models with multimodal experiments by using AI approaches to automate common tasks in multiscale modeling. }
  \label{fig:self-driving}
\end{figure}

Here, we propose the idea of ``self-driving models'' that can automate or accelerate the process of connecting multiscale catalysis models with multimodal catalytic data collected from experiments, as illustrated in Fig. \ref{fig:self-driving}. This idea builds on recent advances in agentic AI for scientific discovery \cite{Xin2025b}, with a specific focus on the opportunities and challenges in catalysis science. Similar to self-driving labs, which automate some or all tasks associated with synthesis, characterization, or testing, a self-driving model is one which can automate some or all tasks associated with constructing, refining, and validating a multiscale catalysis model by comparing it directly to measured kinetic and spectroscopic data. 
The result of such an approach is inherently \textit{interpretable} since the outcome will be well-defined computational models based largely or entirely on known physical or statistical models. These multiscale models can be used to create process models, generalized to similar reactions, and easily shared with the community. In other words, the outcome of a self-driving model is simply a collection of multiscale models that have been automatically constructed and refined using strategies that will vary depending on the specifics of the self-driving model architecture and the underlying catalysis problem. Catalysis science is particularly well-suited for self-driving models because (1) the multiscale models used in catalysis require a large number of assumptions, simplifications and numerical methods, particularly when seeking fundamental insight associated with intrinsic kinetic models \cite{Micale2022}, (2) it is rarely possible to directly observe catalytic reactions, so it is typically necessary to synthesize measured data across many different types of multimodal experiments with complementary strengths and weaknesses, and (3) models are needed to interpret measurements, and measurements are required to calibrate and refine models. For these reasons, quantitative comparisons between multiscale models and experimental observables currently require significant human intelligence and effort, and are relatively rare in the catalysis literature \cite{Sutton2013, Xie2022, Kreitz2023, Lin2025, OConnor2025}.


On the modeling side, a known challenge in using multiscale models for catalysis is that the models have many adjustable, assumed, or estimated parameters. Even if the parameters are computed from first-principles models or fit to experimental data, there are many assumptions required (e.g., type of active site, reaction mechanism, coverage, partition function) and significant uncertainty in the parameter values (e.g., due to underlying DFT approximations or initial guesses in fitting routines). For this reason, the ``inverse problem'' of identifying a fundamental mechanism and an active site directly from data is generally ill-conditioned. 
It is not unusual to have hundreds or thousands of parameters in a kinetic model, and the addition of reactor models, measurement models, and atomistic models could easily require thousands of parameters, leading to a ``many-to-one'' mapping. 
For example, many different sets of kinetic and thermophysical parameters can lead to the same kinetic observables \cite{Bhandari2020, Yonge2024}, and ambiguities in atomic positions and exchange-correlation functionals make it possible for many different atomic-scale models to yield similar rate parameters \cite{Kreitz2023}. The high dimensionality of the problem makes it difficult to find even one set of parameters that quantitatively matches experiments, and the ill-conditioned nature means that there are likely many other possibilities that may not be explored because of limited resources and bias in exploring the high dimensional space. Self-driving models can help overcome both challenges by enabling faster and more thorough searches that keep track of all models explored, giving insight into sensitivity and uncertainty. Using modern advances in computing, ML interatomic potentials, and kinetic modeling, it is possible to envision solving $\sim 10^6 - 10^8$ instances of a ``forward problem'' to identify ensembles of multiscale models that agree with experiments, yielding insight into parameter sensitivity, model robustness, and competing mechanistic hypotheses. However, running complex numerical models at this scale will be difficult or impossible if they require significant human intervention.


Direct comparison with experimental data is also critical for self-driving models. Experimental data in catalysis is often complex to interpret for many reasons. Kinetic information is typically gathered under different reaction conditions (e.g. temperature, pressure, concentration), and catalyst states (e.g. pre/post mortem, \textit{operando}, supported/unsupported). Moreover, the kinetic information is generally convolved with other physical parameters (e.g. surface area, porosity, particle size distribution) and heat or mass transport effects. Most adsorbed intermediate states (and solid active sites) cannot be directly measured, and require multimodal spectroscopic techniques which must be deconvoluted. Measurements are often expensive and time-consuming, and data may be scarce or incomplete (e.g., gas-phase measurements are available at different conditions than spectroscopic measurements of surface adsorbates). Furthermore, there are well-known challenges with data quality in catalysis \cite{Bhan2022, Flaherty2024, Wrasman2024}, which means that even experimentally measured data may be inconsistent or contain ambiguities. An often overlooked aspect in catalysis science and modeling is the role of ``measurement models'', such as the defragmentation of mass spectra, which is known to be challenging for complex species \cite{Beck2024}.
Omitting the measurement model introduces an uncontrolled source of uncertainty in direct comparisons between multiscale models and experimental data. In principle, it is possible to incorporate measurement models into a multiscale modeling framework, but doing so requires even more approximations and parameters, and is rarely done. However, self-driving models can help overcome this barrier by more rapidly exploring this additional layer of model complexity, allowing them to consider the implications of imperfect measurements. 
Self-driving models could also perform tasks such as identifying subsets of a data set that are inconsistent with each other, prompting a deeper critical evaluation of the data or new experimental measurements. Self-driving models could even be incorporated into self-driving laboratories to act as an advanced form of model-based design of experiments or ``knowledge engines'' \cite{Caruthers2003}. Although ambitious, the core capabilities for achieving such a vision largely already exist, and we briefly review them here.

\section{Atomistic, kinetic and reactor modeling}
Recent advances in ML and numerical methods have reshaped the capabilities in multiscale models in catalysis. At the scale of atoms, ML and reactive force fields have made it possible to sample and relax complex active sites, such as supported nanoparticles, with near-DFT accuracy at reduced computational cost \cite{Chanussot2021, Kozinsky2022, Jung2023}. These capabilities have also allowed significant improvement in computing enthalpies and pave the way for statistical sampling techniques to derive more accurate entropies \cite{Stocker2023,Gupta2024,Blondal2024}, enabling the direct calculation of free energies for kinetic models. Microkinetic and kinetic Monte Carlo (kMC) models have also advanced in scope and computational efficiency. kMC models can capture complex effects like site heterogeneity \cite{Deimel2022}, and have been coupled directly to ML force fields \cite{Yokaichiya2024, Mou2022}. Microkinetic models have also advanced, with ML playing a key role in tracking complexities related to adsorbate-adsorbate interactions \cite{SchwalbeKoda2025, Rangarajan2022}.

\begin{figure}[htbp]
  \centering
  \includegraphics[width=0.9\textwidth]{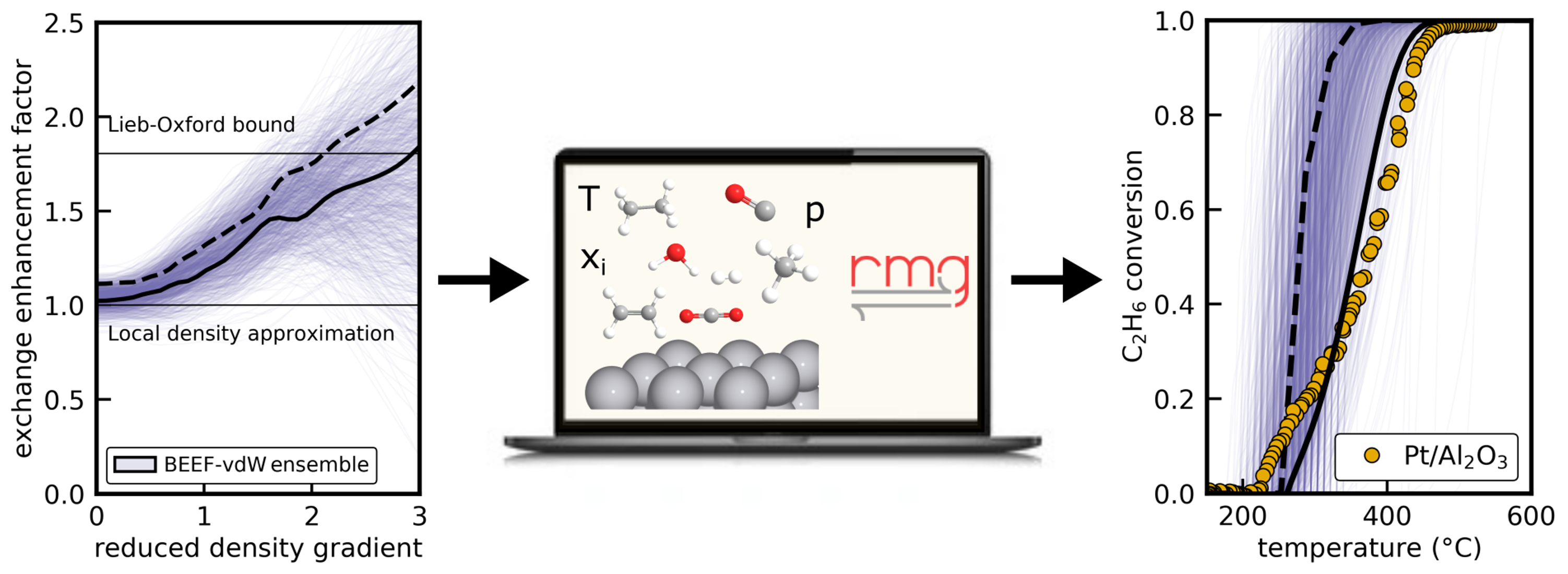}
  \caption{Illustration of results adapted from Kreitz et al. \cite{Kreitz2023}, where an ensemble of exchange enhancement factors from the BEEF-vdW functional was used to propagate uncertainty through adsorption energies, reaction mechanism generation, and reactor modeling to generate an ensemble of light-off profiles for exhaust gas oxidation. The results were directly compared with experiments, allowing identification of the exchange-correlation functional that was most consistent with experimental observations.}
  \label{fig:mkm}
\end{figure}

Beyond these first-principles models, advances in mechanism generation and reactor modeling have expanded the scope of what is possible in comparison between multiscale models and experiments. Automated generation of reaction networks makes it easier to determine the detailed chemical kinetics, and reduces bias through systematic exploration of all possible reactions \cite{Goldsmith2017, Steiner2022}. ML approaches expanded the capabilities of reaction network generation, enabling faster and more accurate determination of complex reaction networks as well as their energetic parameters \cite{Liu2021,Ureel2025}. Moreover, ML and other numerical techniques have expanded capabilities in reactor modeling, enabling the coupling of complex microkinetic models with realistic reactor models for direct prediction of experimentally observable rates \cite{Biermann2025, Hulser2025}. In one noteworthy example, illustrated in Fig. \ref{fig:mkm}, Kreitz and co-workers used an ensemble of automatically generated first-principles multiscale models, derived from the BEEF-vdW ensemble of exchange-correlation functionals, to solve for 2000 different light-off profiles for the oxidation of exhaust gas emissions over a Pt(111) catalyst \cite{Kreitz2023}. For each different set of DFT energies, the mechanism was determined independently with the Reaction Mechanism Generator (RMG), and the resulting ensemble was compared to experiments to quantify uncertainty and reveal the functional that was most consistent with experiment. This example, which took only a few hours on a cluster, reveals how rapidly multiscale models can be solved with modern methods, and illustrates the power of using ensembles of correlated models in comparison with experiments.

Notably, there are several multiscale modeling frameworks designed specifically for catalysis such as the Virtual Kinetics Lab (VLab) \cite{Zong2023}, CATKINAS \cite{Xie2022}, AMUSE \cite{SabadellRendon2023}, Genesys-Cat \cite{Ureel2025}, and RMG \cite{Goldsmith2017,Liu2021}. These toolkits automate many elements of multiscale model construction, but still require significant human effort to install programs, set up models, debug outputs, and refine models. These practical barriers are time-consuming and hinder widespread adoption, but these are also problems that AI and large language models (LLMs) are increasingly proficient at solving. Coupling AI models with these existing frameworks, or using it to piece together new frameworks, is a promising route to creating self-driving models.

\section{Experimental techniques for multimodal data generation}

\begin{figure}[htbp]
  \centering
  \includegraphics[width=0.9\textwidth]{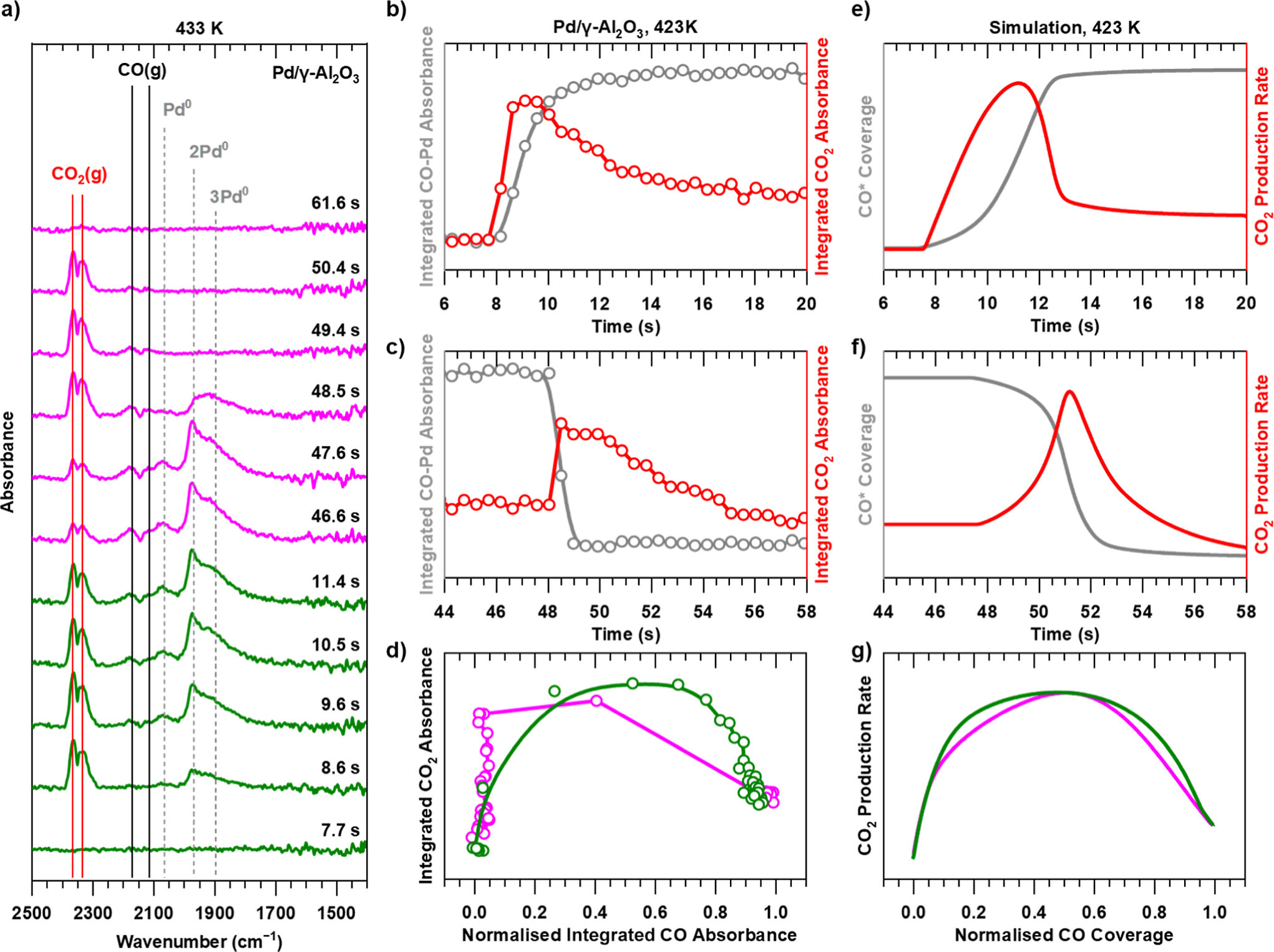}
  \caption{(a) Time-resolved DRIFTS spectra for \ce{CO2} production over Pd/$\gamma$-\ce{Al2O3} catalyst with time-resolved intensities of integrated CO* peaks for the (b) forward and (c) backwards transients, and (d) comparison of \ce{CO} and \ce{CO2} absorbance during the forward (green) and backwards (magenta) transients. Direct comparison to simulated coverages and rates from microkinetic models are shown in panels (e-g). Reproduced with permission from Ref. \citenum{OConnor2025}. }
  \label{fig:expt}
\end{figure}

Fundamental catalysis data, like steady-state conversion/rates and kinetic observables (e.g. apparent activation energies, reaction orders and transfer coefficients), form the foundation of mechanistic investigations in catalysis \cite{lynggaard_analysis_2004, campbell_analysis_2021}, but modern advances in \textit{operando} spectroscopy and transient kinetic techniques now generate multimodal datasets that reveal unprecedented detail about catalytic chemistry, capturing both the evolving structure of active phases and the dynamics of adsorbed intermediates under reaction conditions. \textit{Operando} methods such as X-ray absorption, infrared, and Raman spectroscopy enable real-time observation of transient species and surface transformations that cannot be accessed ex situ \cite{Weckhuysen2003, Hermans2025}. Transient approaches such as temporal analysis of products (TAP) \cite{Fushimi2025, OConnor2025}, steady-state isotopic transient kinetic analysis (SSITKA) \cite{Janssens2023}, and temperature programmed surface reaction (TPSR) \cite{Reece2021} can probe adsorption, reaction, and desorption kinetics by monitoring system responses to controlled perturbations. Recent studies highlight how TAP elucidates kinetics under pulsing conditions \cite{Fushimi2025, OConnor2025}, SSITKA quantifies surface residence times and links them to microkinetic models \cite{Janssens2023}, and TPSR reveals thermally activated surface processes \cite{Reece2021}. Coupling transient approaches with kinetics can correlate structural signals with transient responses, enabling the separation of active intermediates from spectator species and directly connecting spectral features to kinetics \cite{Hermans2025, Reece2021, OConnor2025}. For example, Reece and co-workers recently demonstrated semi-quantitative agreement between transient diffuse reflectance infrared Fourier transform spectroscopy (DRIFTS) signals and microkinetic models for CO oxidation on a supported Pd/$\gamma$-\ce{Al2O3} catalyst, as illustrated in Fig. \ref{fig:expt}.  While these techniques provide a wealth of mechanistic information, the richness and heterogeneity of the resulting multimodal datasets make them challenging to analyze quantitatively, limiting the ease with which they can be compared across studies or incorporated into predictive models.

A major barrier to broader integration of catalysis data with machine learning and AI is the uncertainty and lack of standardization in experimental data and associated measurement models in catalysis. Unlike thermodynamic data, which are widely tabulated and standardized, transient kinetic and \textit{operando} spectroscopic measurements are highly method-specific and often lack uniform reporting conventions. This heterogeneity makes it difficult to compare datasets across research groups or to build consistent repositories for benchmarking theory. However, recent efforts to standardize and tabulate catalytic data have emerged \cite{Burte2025, Huskova2025, Marshall2023}. Although these studies represent a significant step forward, most efforts have focused on steady-state measurements and materials characterization data, while more complex datasets from transient and spectrokinetic measurements are less standardized and the raw data is often not provided. The inherently heterogeneous nature of data in catalysis \cite{Marshall2023}, and the particular complexities of transient and \textit{operando} measurements, makes it unlikely that a single data standard will ever take over. However, the possibility of self-driving models that incorporate measurement models (i.e. the pre-processing is included as part of the model) may reduce some of the burden associated with data standardization by working directly with raw data. Moreover, self-driving models can be constructed to work with heterogeneous and complex datasets such as transient spectra measured at different conditions or even in different studies, potentially identifying discrepancies and extracting new insights from these time-consuming and expensive measurements. These opportunities can help incentivize data sharing and standardization in catalysis science.

\section{Model fitting, generative models, and agentic AI}

\begin{figure}[htbp]
  \centering
  \includegraphics[width=0.8\textwidth]{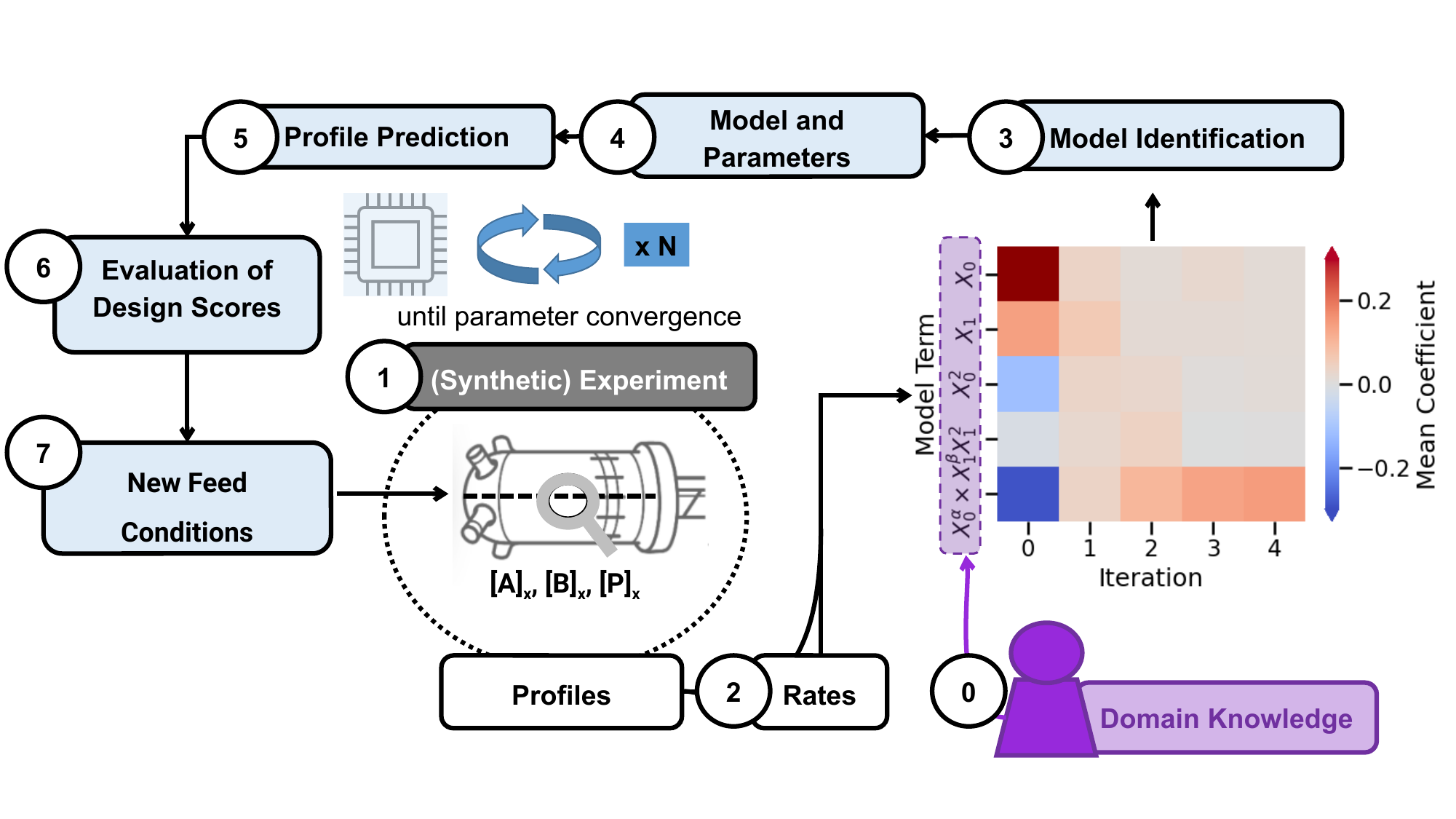}
  \caption{Illustration of model-driven adaptive design of experiments using a synthetic profile reactor setup. The architecture uses a feedback loop to identify models and parameters and suggest new experiments. Reproduced with permission from Ref. \citenum{Kouyate2025}.}
  \label{fig:fitting}
\end{figure}

The analysis and design of catalytic systems can be cast as an inverse problem, in which the desired outcome (e.g., activity, selectivity, or stability under operating conditions) is specified, and the goal is to determine the catalyst structure or reaction conditions that will produce it. Such formulations have a long tradition in process systems engineering, where optimization frameworks are used to identify kinetic models from data and thermodynamic constraints, although there are well-known challenges with multiple local minima and insensitive parameters \cite{Bhandari2020, Chacko2023, Lejarza2023}.  In a particularly relevant recent paper, Kouyate et al.~\cite{Kouyate2025} construct an adaptive model for automatically optimizing a kinetic model to a profile reactor using optimization and ML \cite{Kouyate2025}, as illustrated in Fig. \ref{fig:fitting}. 

In parallel with advances in optimization, there has been growing interest in applying generative ML models to inverse problems in chemical engineering. Recently, Alves and Kitchin~\cite{Alves2025} have explored the possibility of using generative models for optimization. This perspective shows how generative models can directly confront the ``many-to-one'' issue that plagues many other forms of optimization. Generating samples from a learned distribution provides a way to find multiple solutions (or initial guesses), reducing challenges with local minima without the prohibitive cost of global optimization in high-dimensional problems. Self-driving models combine this paradigm with generative AI models that can automatically ``generate'' multiscale models. In the short term, this will likely be achieved by coupling of generative language models with existing multiscale modeling tools, but in the long term it is possible to envision specialized generative models for catalysis that can directly generate multiscale models or initial guesses of rate pamareters from kinetic data.

\begin{figure}[htbp]
  \centering
  \includegraphics[width=0.8\textwidth]{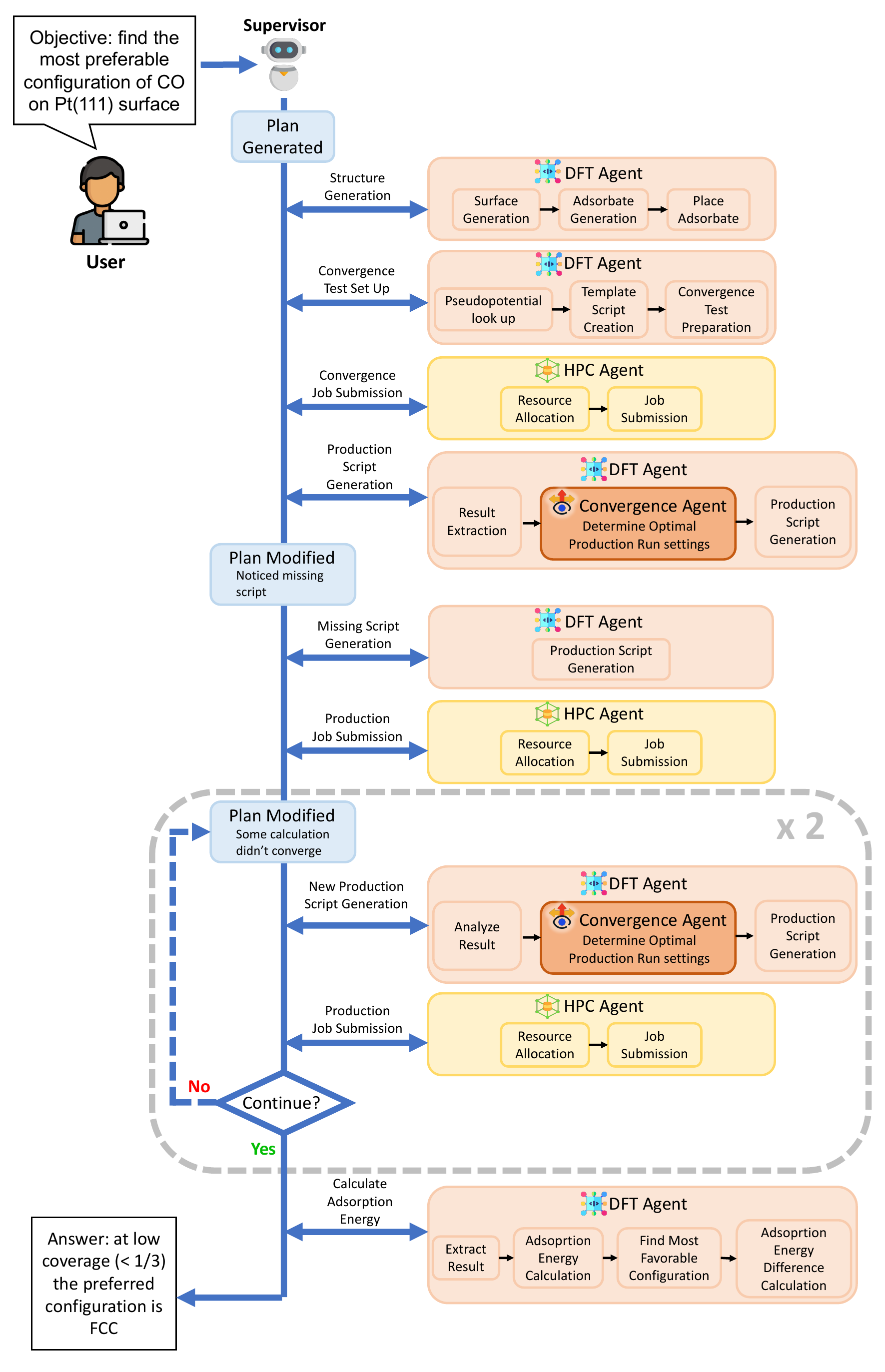}
  \caption{Workflow for CO/Pt(111) study done by an automated agentic system for setting up, running, converging, and evaluating DFT simulations of surfaces and adsorbates.}
  \label{fig:DREAMS}
\end{figure}

One strategy for creating self-driving models is to leverage the rapidly maturing toolkit of LLMs and agentic AI systems \cite{Xin2025b}. Pre-trained LLMs are becoming ubiquitous, and can be customized by fine-tuning them with domain data, incorporating in-context learning, or using specialized templates and retrieval-augmented generation (RAG) techniques \cite{Porosoff2025, DREAMS2025, MendibleBarreto2025, Campbell2025}. For example, recent work by White and co-workers~\cite{Porosoff2025} provides an example of how in-context learning with pre-trained LLMs can drive catalyst discovery. Another salient example is the work by Viswanathan and co-workers~\cite{DREAMS2025}, who demonstrated an agentic system for running and validating DFT simulations of various properties, including adsorption energies. This example, illustrated in Fig. \ref{fig:DREAMS}, is a particularly relevant demonstration of the capabilities of agentic systems in automatic complex computational pipelines that are directly relevant to catalysis. Similar strategies could be used to create agents that can autonomously run software to explore active sites structures with ML force fields, generate reaction mechanisms, or solve reactor models. These agents could be combined with straightforward logic to automatically generate large collections of multiscale model results. Similar agents could be used to compare results with experimental data, sample possible inputs, run models, and adjust inputs to improve agreement with measured multi-modal data.  Development of specialized generative models for specific tasks like active site \cite{Kolluru2024} or mechanism generation could further accelerate searches. A natural extension of these approaches lies in model refinement, driven by AI models, optimization models, or combinations of the two. For example,  model-based design of experiments \cite{Yonge2024, Franceschini2008} provides a route to systematically identify experiments that refine the precision of parameters or differentiate between multiple possible models. If AI systems can accelerate the generation of ensembles of models that fit the data, then these optimization techniques can help identify critical experiments that most effectively discriminate among competing mechanistic hypotheses or reduce the uncertainty on a model parameter. Alternatively, existing agentic AI systems for scientific discovery \cite{SciToolAgent} or self-driving labs \cite{Tom2024} may enable more effective model refinement or experimental planning.

\section{Conclusions and Outlook}
\label{sec:conclusions}

The vision of self-driving models extends the paradigm of self-driving labs to the computational domain, aiming to automate the construction, refinement, and validation of multiscale catalytic models through direct comparison with multimodal experimental data. Advances in atomistic simulation, ML force fields, microkinetic and reactor modeling, operando spectroscopy, and transient kinetics have created both the rich data and the computational tools needed to make this vision feasible. Generative AI, particularly when combined with optimization frameworks and agentic AI systems, provides a natural way to drastically scale up the construction, refinement, and analysis of multiscale models. Increasing the speed and scale of multiscale modeling has the potential to open up qualitatively new approaches to address the ``many-to-one'' nature of the inverse problem in catalysis by generating large ensembles of plausible mechanisms and models, while model-based design of experiments offers a pathway to systematically reduce uncertainty. Although significant challenges remain, the core components are already in place. By linking mechanistic reasoning with automated workflows, self-driving models promise not only to accelerate understanding of intrinsic kinetics from measured data, but also to enhance reproducibility, broaden mechanistic exploration, and reduce bias in catalytic science. In the long term, these systems may serve as community knowledge engines, continuously integrating new data and models to deliver interpretable, transferable, and predictive frameworks for catalysis and other related application domains where complex models and multimodal data need to be synthesized to gain scientific insight.


\section{Acknowledgements}
A.J.M. and T.N.W. acknowledge funding from the National Science Foundation under award number  1943707.

\bibliographystyle{elsarticle-num} 
\bibliography{cas-refs}

\vspace{12pt}
References and recommended reading. Papers of particular interest, published within the period of review, have been highlighted as:

* of special interest

** of outstanding interest
\end{document}